# Ultrafast Optical Generation of Coherent Bright and Dark Surface Phonon Polaritons in Nanowires

Pierre-Adrien Mante,[1, 2, *] Sebastian Lehmann,[3] Daniel Finkelstein Shapiro,[1] Kayin Lee,[2] Jesper Wallentin,[4] Magnus T. Borgström,[3] and Arkady Yartsev[1]

[1] Division of Chemical Physics and NanoLund, Lund University, Sweden.

[2] Department of Applied Physics, Hong Kong Polytechnic University, Hong Kong S.A.R..

[3] Division of Solid-State Physics and NanoLund, Lund University, Sweden.

[4] Division of Synchrotron Radiation Researchand NanoLund, Lund University, Sweden.

E-mail:* pierre-adrien.mante@chemphys.lu.se

**Abstract:**

The sub-wavelength confinement and enhanced electric field created by plasmons allow precise sensing and enhanced light-matter interaction. However, the high frequency and short lifetime of plasmons limit the full potential of this technology. It is crucial to find substitutes and to study their dynamics. Here, we propose an experimental approach allowing the time-domain study of surface phonon polaritons. We first build a theoretical framework for the interaction of ultrashort pulses of light with polar materials. We then perform femtosecond pump-probe experiments and demonstrate the generation and time-resolved detection of surface phonon polaritons. By comparing experiments and simulations, we show the generation of both bright and dark modes with quality factor up to 115. We then investigate mode dependent decay and energy transfer to the environment. Our results offer a platform for the experimental exploration of the dynamics of surface phonon polaritons and of the role of coherence in energy transfer.

Plasmonics enables numerous applications thanks to the deep sub-wavelength confinement of electromagnetic waves.[1-11] The resulting enhanced electromagnetic fields have been used to achieve precise sensing, both through the sensitivity to refractive index changes[2] and the associated enhancement of nonlinear optical effects, such as surface-enhanced Raman scattering (SERS).[3] Due to the confinement of the electric field, plasmonic structures can also replace optical cavities.[4-8] For instance, when a molecule and a plasmonic structure are close to each other, excitons and plasmons strongly couple to form quasi-particles with hybrid properties: plexcitons.[4-6] While the exceptional spatial confinement enabled by plasmons offers countless prospects, certain characteristics have become a hurdle that reduces their field of applications. First, plasmons suffer from large losses due to Landau and radiative damping.[9] These losses broaden plasmon resonances and reduce their sensing capabilities.[10] Besides, plasmons in metallic structures have high frequency, and the direct study of the phase of coherent plasmons in the time domain requires advanced spectroscopic techniques technologies.[11] Organizing plasmonic particles in an array gives rise to surface lattice resonance that have long lifetime, but strong spatial confinement disappears.[12] To achieve the full potential of sub-wavelength confinement and to better understand energy transfer, it is crucial to find alternatives to plasmons with sub-wavelength confinement and long lifetime and to study their dynamic.

Surface phonon polaritons (SPhPs) results from the coupling of light in the THz to mid-infrared frequency range to phonons in polar materials and offer similar prospects to plasmons.[10, 13-17] Various aspects of SPhPs have been studied, from radiative heat transfer[13, 14] to ultra-confinement of light at THz frequencies,[15-17] and their coupling to carrier plasma.[18] Like plasmons, SPhPs allow deep sub-wavelength confinement. Moreover, SPhPs have a much longer lifetime and sharper resonances, which makes them more sensitive to change in their environment,[10] but also able to achieve large Purcell factor ($10^6$-$10^7$).[15] SPhPs also allow enhanced electric field, which can strongly couple to excitation at the same energy, like molecular vibrations[19] and propagating SPhPs in a nanowire (NW) array.[20] Finally, SPhPs have relatively low frequency and long lifetime, which should permit their time-resolved study using current commercial technologies, such as femtosecond lasers.

In this letter, we study the generation and time-resolved detection of SPhPs in InP NWs using femtosecond pump-probe spectroscopy. We first review the different mechanisms for the generation and detection of coherent vibrations by light pulses in polar semiconductors in the presence of a surface field. We then perform experiments on InP NWs with various doping concentrations and verify the generation of SPhPs. By combining experiments and simulations, we study various aspects of SPhPs, from their sensing potential to their lifetime and dissipation. Our work offers a novel platform for the study of the dynamics of processes involving SPhPs, from radiative heat transfer[13] to strong coupling.[19, 20] This experimental approach will also permit the study of the rich physics of polaritons in the time domain, including the manipulation of quantum information and coherent energy transfer.[6, 21-23]

## Results

### Theoretical principle

Infrared absorption spectroscopy on nanostructures is the most common experimental method to study SPhPs.[15, 20] Unfortunately, this steady-state approach does not allow the retrieval of the coherent dynamics of SPhPs. To do so, we must generate a coherent population of SPhPs and sample the evolution of its amplitude and phase. Since phonons have frequencies of few THz (periods of hundreds of fs), femtosecond laser pulses are the obvious choice. However, visible pulses have energy much higher than SPhPs and cannot couple directly due to energy and momentum mismatch. However non-linear and current-induced effects can be harnessed to create a polarization at THz frequencies in a material (Fig. 1), which acts as a driving force for SPhPs (see Supplementary note 1). Such methods have been extensively used to generated THz electromagnetic waves and coherent phonons in polar materials.[24-29]

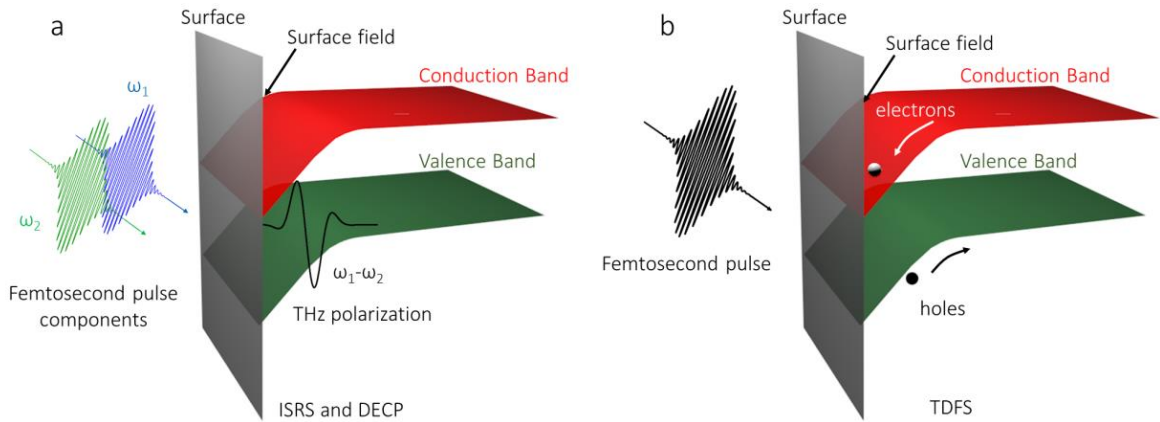

Fig. 1. Generating surface phonon polaritons. The phonons are generated through the creation of a polarization at THz frequencies that acts as a driving force. Multiple mechanisms can create such polarization. (a) Schematic representation of the impulsive stimulated Raman scattering (ISRS) and the displacive excitation of coherent phonons (DECP) mechanisms responsible for the generation of surface phonon polaritons. (b) Schematic representation of the transient depletion field screening (TDFS) mechanism responsible for the generation of surface phonon polaritons.

The absorption of a femtosecond laser pulse in a polar material creates coherent vibrations at THz frequencies. [24-29] The principal mechanisms for such generation are impulsive stimulated Raman scattering (ISRS),[24] displacive excitation of coherent phonons (DECP),[25] and transient depletion field screening (TDFS).[26, 27] In ISRS and DECP, mixing of different frequency components of the femtosecond pulse creates a polarization at the frequency of the vibration (Fig. 1a).[28] ISRS and DECP correspond to non-resonant and resonant processes, respectively and both are second-order optical nonlinear effects described by the susceptibility $\chi^{(2)}$.[28] In addition to these mechanisms, third-order effects arise in the presence of a surface field, $E_s$. The third-order susceptibility, $\chi^{(3)}$, characterizes such processes. In TDFS, the generation of photocarriers leads to charge currents, *J*, and to the set-up of a polarization in the medium. Currents are created by the presence of a surface field that displaces photo-excited carriers (TDFS, Fig. 1b),[27] or by the ambipolar diffusion of charges (photo-Dember effect).[29] The current created by photoexcited carriers in a surface field is directly proportional to the amplitude of the field. Considering all these effects, the coherent phonon amplitude, *Q*(t), is:

$$Q(t) \propto -\frac{e^*}{\varepsilon_0 \varepsilon_\infty}\left(\chi^{(2)}I_{pump} + \chi^{(3)}I_{pump}E_s + \int_{-\infty}^{t} J(t')dt'\right)$$

$$= -\frac{e^* I_{pump}}{\varepsilon_0 \varepsilon_\infty}[\chi^{(2)} + E_s(\chi^{(3)} + \alpha)] \quad (1)$$

where *e\**, is the effective lattice charge, $\varepsilon_0$ and $\varepsilon_\infty$, are the static and high frequency dielectric constant, respectively, $I_{pump}$ is the laser intensity and α is a proportionality coefficient that consider the dynamic of charges in the surface field and the material properties. A detailed presentation of the generation mechanisms and the role of polarization can be found in Supplementary note 1.

For the detection of coherent vibrations with femtosecond light pulses, nonlinear effects also play a fundamental role. Anti-Stokes Raman scattering of an incoming photon produces a shift in its frequency. As was the case for the generation, the susceptibilities tensors, $\chi^{(2)}$ and $\chi^{(3)}$, rule these processes. The scattered photons then interfere with the undisturbed photons of the femtosecond pulse to provide a heterodyne detection, which allows the retrieval of the phase of the coherent vibrations (see Supplementary Note 2). Overall, we can write the change of reflectivity induced by the vibration *Q(t)* as:

$$\Delta R(t) \propto I_{pump}[\chi^{(2)} + E_s(\chi^{(3)} + \alpha)] \cdot I_{probe}(\chi^{(2)} + \chi^{(3)}E_s) \quad (2)$$

where $I_{probe}$ is the intensity of the probe laser. From Eqs. (1) and (2), we see that the surface field, $E_S$, influences the amplitude of the signal. Since the amplitude of the surface field is proportional to the square root of the doping concentration, $N_D$,[30] it offers an approach to isolate the generation and detection mechanisms. By modifying the density of free carriers, we can control the Fermi level within the nanowire and modify the surface field. When the femtosecond pulse is absorbed, the current created by photo-excited carriers and the amplitude of third order non-linear processes will be proportional to the surface field.

**Generation mechanism**

To verify the possibility to generate SPhPs in NWS made of polar semiconductors, we have grown InP NWs with different doping concentrations ranging nominally intrinsic to $N_D = 4.5 \times 10^{19}$ cm$^{-3}$ (see Methods for detailed sample growth process). Figure 2 shows a 30° tilted scanning electron micrograph of the NWs sample with a doping concentration of $N_D = 1.28 \times 10^{19}$ cm$^{-3}$. We use the Burstein-Moss shift of the photoluminescence to obtain the doping concentration at room temperature in each sample. Further details on this procedure, as well as additional characterizations of these specific NWs samples can be found in Ref. 31. The average diameter of the NWs diminishes when increasing sulfur doping and range from 124 ± 12 to 92 ± 8 nm. For each sample, the diameter dispersion is roughly 10 %. The height ranges from 1100 ± 40 to 2800 ± 70 nm.[31] The diameter and height distributions are obtained by performing the statistical average above a large number of measurements on SEM images. The NWs exhibit Fermi level pinning, and for the doping concentrations we consider, the depletion layer thickness is smaller than 20 nm according to a combined X-ray photoemission electron and scanning tunneling microscopy study.[32]

We performed ultrafast pump-probe spectroscopy on the as-grown free-standing NWs samples as shown in Fig. 2 (see Methods for additional details on the experimental setup). A femtosecond laser with repetition rate of 2 kHz delivers pulses at 1030 nm and 200 fs duration that are used to pump two Non-collinear Optical Parametric Amplifiers (NOPAs). From the first NOPA, we obtain pulses at a wavelength of 550 nm that are compressed down to 35 fs. These pulses are used as pump since they have energy larger than the bandgap of InP and are thus able to generate carriers. The second NOPA delivers 720 nm pulses with 40 fs duration. These pulses are sent through a delay stage, in order to control the time between the arrival of the pump and the probe pulses. The pump beam modulated at a frequency half the repetition rate of the laser, thus enabling the measurement of the probe reflectivity with and without the effect of the pump, $R_{\text{pump}}$, and $R_0$, respectively. The transient reflectivity is then obtained by calculating $(R_{pump} - R_0)/R_0$. In order to study the effect of the polarization of the pump and probe, experiments are performed at normal incidence (equivalent to a polarization normal to the NWs axis) or at grazing angle incidence (the polarization can be changed from normal to the NWs axis to almost parallel). The inset of Fig, 3a displays the signal obtained on the NWs with a doping concentration $N_D=2.5\times10^{19}$ cm$^{-3}$.

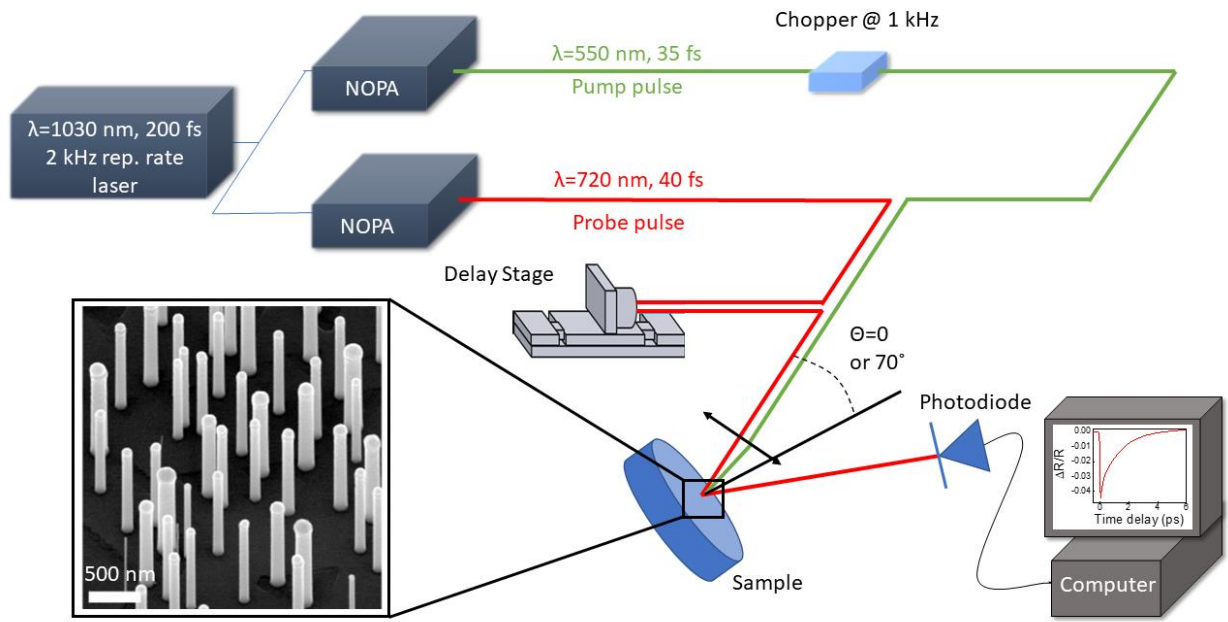

Fig. 2. Schematic representation of the experimental pump-probe setup and scanning electron micrograph of the sample. Experiments are performed at normal incidence or at grazing angle incidence.

To obtain the coherent oscillations (Fig. 3a), we subtracted the background, which corresponds to the electronic and thermal responses of the sample, using a three-exponential function. The obtained signal has a complex structure with beatings that highlight the presence of multiple frequencies. By performing the Fourier transform of the signal (Fig. 3b), we observe four frequencies at 8.9, 9.1, 9.8 and 10.1 THz. We also measured the Raman spectrum of these InP nanowires (Fig. 3b). In these spectra, we observe only two peaks at 8.9 and 10.3 THz, which we attribute to the TO and LO phonons of InP, respectively. These values agree with values reported in InP nanowires and display a red shift compared to bulk InP.[33, 34] The TO mode is common to both experimental techniques. However, the pump-probe data displays three prominent peaks between the TO and LO phonons frequencies. Due to their location within the Reststrahlen band, a frequency region in which light cannot propagate within a medium and that occurs for polar materials between the TO and LO frequencies, we attribute these vibrations to SPhPs.[16] In the following, we refer to the mode at 9.1 THz as the longitudinal dipole mode (LD), and the modes at 9.8 and 10.1 THz as the transverse quadrupole (TQ) and transverse dipole (TD) modes, respectively. We will explain this nomenclature in the following. Given the high doping in the investigated nanowires, the formation of coupled plasmon-phonon modes could be expected.[27, 35] These modes have signatures, such as the dependence of their frequency on the density of free carriers, that are not observed in our experiments and can thus be ruled out (see Supplementary note 1 and Fig. S1).

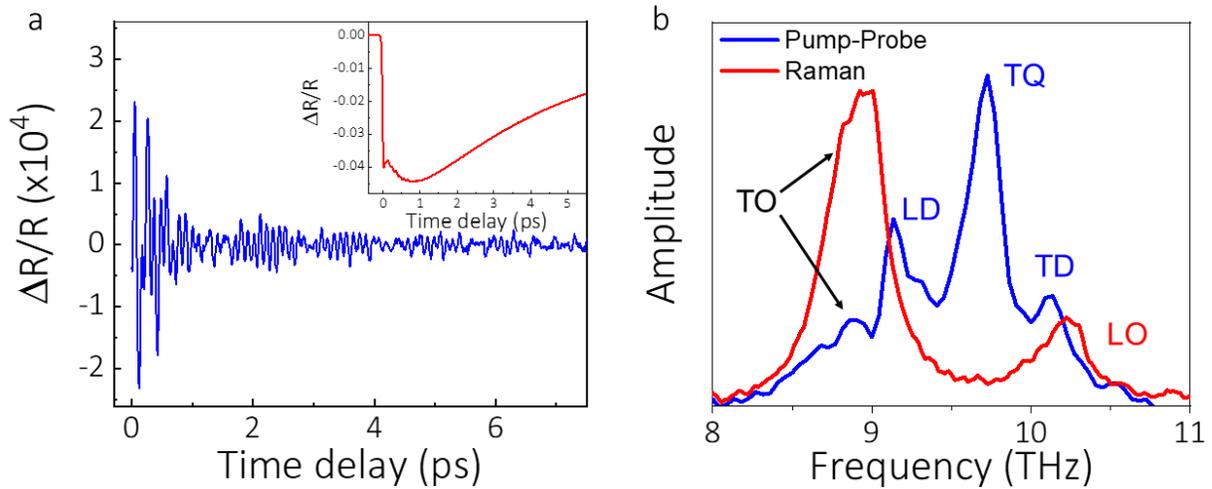

Fig. 3. Pump-probe and Raman experiments. (a) Oscillatory component of the transient reflectivity obtained on InP NWs with a doping concentration ND=2.5×10$^{19}$ cm$^{−3}$ with a pump and probe wavelength of 550 and 720 nm, respectively. Inset: transient reflectivity before subtraction of the electronic and thermal contributions. (b) Fourier transform of the transient reflectivity (Blue line) and Raman spectrum (Red line) obtained on the InP NWs with a doping concentration ND=2.5×10$^{19}$ cm$^{−3}$. The Raman spectrum shows the transverse (TO) and longitudinal (LO) optical phonons of InP, while the pump-probe data shows the transverse optical phonons (TO) of InP, as well as the longitudinal dipole (LD), transverse dipole (TD) and transverse quadrupole (TQ) surface phonon polaritons.

We then investigate the generation and detection mechanisms of the SPhPs. To do so, we extracted the amplitude of the vibrations as a function of the doping concentration. In Fig. 4a, we show the Fourier transform of the transient reflectivity obtained on samples with different doping concentrations ranging from nominally intrinsic to $N_D$= 4.5×10$^{19}$ cm$^{−3}$.

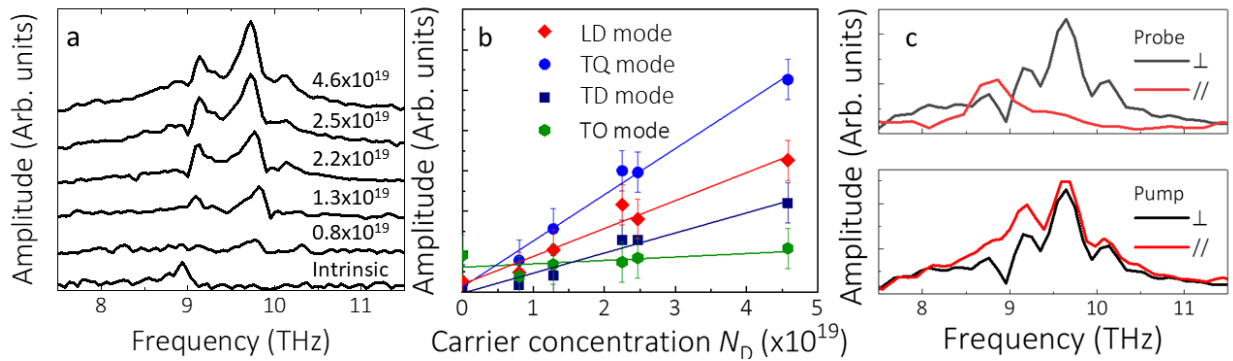

Fig. 4. Generation and detection mechanisms. (a) Fourier transforms of the transient reflectivity obtained on nanowires with various doping concentrations. The amplitude of the signal increases for increasing doping. (b) Amplitude of each experimentally observed mode as a function of doping concentration. (c) Upper panel: Fourier transform of the transient reflectivity obtained for the pump polarization perpendicular to the nanowires axis and the probe polarization either parallel or perpendicular to the nanowires axis. Lower panel: Fourier transform of the transient reflectivity obtained for the probe polarization perpendicular to the nanowires axis and the pump polarization either parallel or perpendicular to the nanowires axis.

We see the amplitudes of the SPhPs steadily increasing with the doping concentration. From this result, we conclude that the surface field is involved in their observation. In Fig. 4b, we show the amplitude of the various peaks in the Fourier transform as a function of the doping concentration. We fit these data with linear functions, and two trends emerge. First, the TO mode does not display any doping concentration dependence. We can conclude, according to Eq. 2, that this mode is generated and detected through second-order processes. Then, the amplitudes of the three modes located in the Reststrahlen band increase linearly with doping concentration. Since the surface field is proportional to the square root of the doping concentration, we can conclude that both the generation and detection mechanisms involve the surface field (See Eq. 2). The generation of these modes can either occur through a third-order process or through TDFS, while for the detection, the linear dependence implies that the responsible mechanism is the third-order frequency mixing process involving the surface field. This detection mechanism is like the Franz-Keldysch effect.[36]

We can further separate the generation mechanism using polarization-dependent experiments (Fig. 4c). These experiments were performed at grazing angle incidence as indicated in Fig. 2. With this setup we can vary the polarization from perpendicular to the nanowires to almost parallel to the nanowires. The frequencies obtained with the pump and probe polarization perpendicular to the nanowires for normal and grazing angle incidence are identical (Fig. S2), which confirms that this is a polarization effect. Mechanisms involving charges, such as TDFS, should be insensitive to the polarization of the pump in contrast to ISRS and DECP. We performed experiments with the polarization of the pump either perpendicular or parallel to the nanowires. We did not observe a significant change in the amplitude of the signal. From this observation, we conclude that the TDFS mechanism is responsible for the generation of the SPhPs, and not nonlinear effects that are highly polarization sensitive.[27, 37] However, experiments with the probe polarization perpendicular and parallel to the nanowires (Fig. 4c) show variations in the amplitude of the signal, as expected from nonlinear effects (see supplementary note 2).

**Surface phonon polaritons**

Now that we have identified the generation and detection mechanisms, we focus on the study of the nature and characteristics of the SPhPs. Like surface plasmons, SPhPs are sensitive to the dielectric function of the surrounding medium.[38] When the dielectric properties change, we expect the frequency of the SPhPs to shift. We thus performed experiments in which we immersed the nanowires in a solution of $CH_2Cl_2$. We chose $CH_2Cl_2$, since it was responsible for the largest shift of SPhPs frequencies in previous studies.[35] In the upper panel of Fig. 5a, we reproduce the Fourier transform of the transient reflectivity obtained in air and $CH_2Cl_2$.

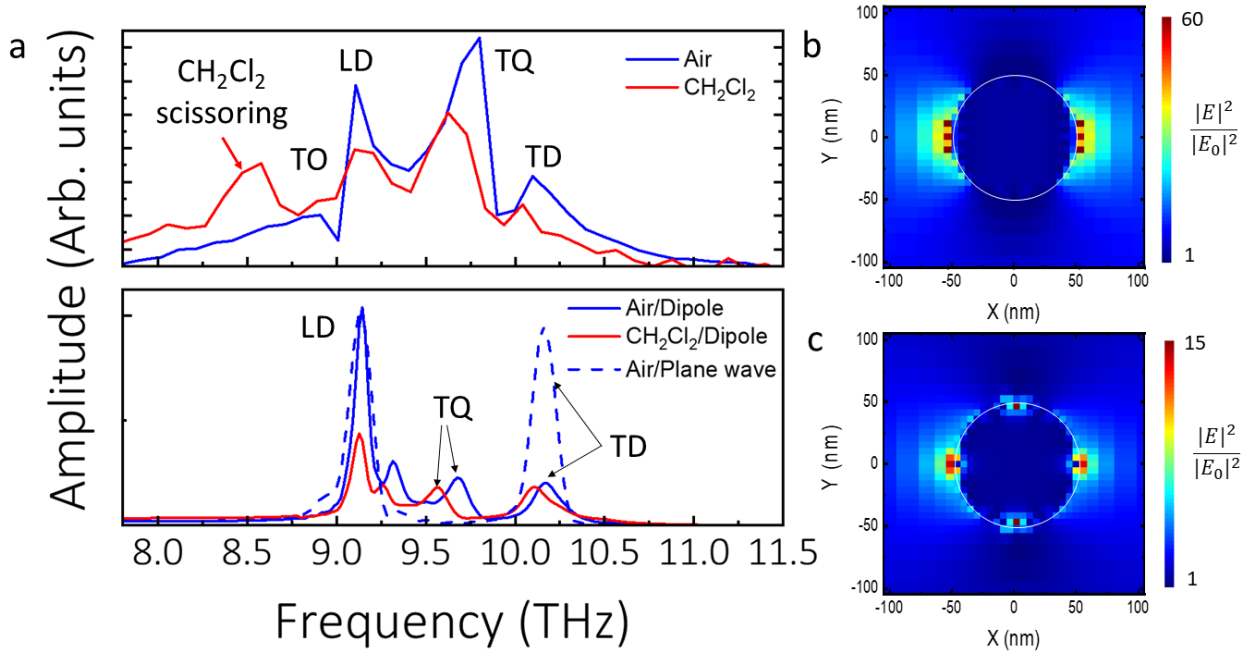

Fig. 5. Surface phonon polaritons modes. (a) Upper panel: Fourier transform of the transient reflectivity signals obtained on InP nanowires in air and in $CH_2Cl_2$. Lower panel: simulation of the absorption by an InP nanowire at THz frequency for an incident plane wave in air (dashed blue line), a dipole located in the vicinity of the nanowire in air (solid blue line), and for a dipole in $CH_2Cl_2$ (solid red line). (b) Electric field distribution in the cross-section of the nanowire at the frequency of the TD mode. (c) Electric field distribution in the cross-section of the nanowire at the frequency of the TQ mode.

In $CH_2Cl_2$, we observe a shift of all the modes except the TO mode, confirming that the LD, TD, and TQ modes are SPhPs. In addition, we observe the appearance of a mode at 8.5 THz, which correspond to the Cl-C-Cl in-plane scissoring.[39] We also confirm that the TD and LO modes are different since the LO mode does not depend on the dielectric properties of the surrounding medium.[40, 41]

To further analyse the different modes, we performed FDTD simulations (see Methods). However, in order to compare these simulations to our experimental data, we need to consider the specificity of our experiments. The electromagnetic modes of a cylinder are classified by using the angular quantum number ($m$) characteristic of the azimuthal mode pattern.[42] This quantum number dictates which mode can couple to an incoming plane wave. For instance, the totally symmetric fundamental breathing mode of a cylinder ($m=0$) cannot be excited using a linearly polarized plane wave at THz frequencies. But for the dipole mode ($m=1$), the symmetry of a linearly polarized wave is similar, and the light can efficiently couple to such mode. In our experiments, in which we are using visible femtosecond laser, we are not exciting SPhPs through the absorption of THz waves, but by using charges as intermediate. This mechanism lifts some of the limitations due to symmetry, which means that we should be able to generate both modes that radiate in the far-field (bright modes), like the $m=1$ mode, and also modes that are confined to the near field due to symmetry reasons (dark modes). Our generation mechanism is thus similar to a near-field excitation. To verify this assumption, we performed FDTD simulations with different sources (see Methods).

We first performed simulations by placing a dipole radiating in the far-infrared region close to the nanowire. This configuration reproduces the conditions for a near field excitation and generates bright and dark modes.[43] The bottom of Fig. 5a shows the scattered spectrum for such a source. We observe a good agreement between experiments and simulation. We then investigate the electric field distribution of each of these modes to be able to identify them. The mode at 9.1 THz (Fig. S3) is the longitudinal dipole of the nanowire, which display extrema at the top and the bottom of the substrate. This mode is called the monopole mode.[15, 44, 45] The modes at 9.8 (TQ) and 10.1 THz (TD) also have maxima at the bottom and the top of the wire but display more complex patterns in the cross-section of the wire (Fig. 5b and c). The mode at 10.1 THz (TD) has two maxima that are diametrically opposed and corresponds to the transverse dipole with $m=1$. Finally, the mode at 9.8 THz (TQ) shows four maxima in the cross-section (Fig. 5c) and corresponds to the transverse quadrupole mode. Given the symmetry of this mode, a plane wave should not be able to couple this mode and it is thus a dark mode that cannot radiate in the far-field. We performed these simulations for the different NWs we investigated experimentally (Fig. S4). We did not observe any significant changes of the frequencies as expected since the dimensions of the NWs are at least ten times smaller than the equivalent wavelength in vacuum.

To further confirm the bright or dark nature of these modes, we performed experiments, where in place of a dipole, we use a linearly polarized plane wave to excite the nanowire. The bottom part of Fig. 5a shows the scattered spectrum in these conditions. With this scheme, only bright modes can be excited, which are modes that radiatively decay. We conclude that LD and TD modes are bright modes while TQ is a dark mode.

Finally, we simulated the spectrum in the case of excitation by a dipole when the nanowires are in a medium with a refractive index of 1.3,[46] which corresponds to the refractive index of $CH_2Cl_2$ in the THz range. We observe a shift of the modes we have attributed to the SPhPs in excellent agreement with the experimentally observed one. Since in these simulations, we only modify the refractive index of the surrounding medium, we are not able to observe the $CH_2Cl_2$ scissoring mode, and we can conclude that the frequency shift we observe experimentally is solely due to the change of refractive index. This experiment highlights the potential of SPhPs to perform refractive index sensing in the THz frequency range.

**Discussion:**

We have generated coherent bright and dark SPhPs at around 10 THz (30 μm wavelength in air) in approximately 100 nm diameter nanowires, thus achieving extreme confinement of electromagnetic waves. We now take advantage of our time-domain approach to extract the lifetime of SPhPs. We fit the transient reflectivity signal obtained on the sample with doping concentration $N_D=2.5\times10^{19}$ cm$^{-3}$ with three damped sines (Fig. S5). We extract a coherence lifetime in air of 4.02 ± 0.13 ps, 2.20 ± 0.03 ps and 0.81 ± 0.018 ps for LD, TQ, and TD, respectively. These values correspond to quality-factors of 115, 67, and 25, respectively. These values must be compared to what is achieved in noble metal, where the quality-factor of plasmon resonances reaches at best 20-40.[16] Due to the strong dispersion in the geometry of the nanowires, these values are lower bounds of the coherence lifetime and higher values should be achievable. However, we already observe a 4-fold increase compared to the best quality factor for plasmon resonance. If we compare the two transverse modes, which suffer from similar inhomogeneous broadening due to the dispersion in diameter, we can see that TQ, which is a dark mode and cannot decay radiatively, has a lifetime three times larger than TD. The combination of strong confinement and high quality-factor will enable reaching record values of the Purcell factor. Furthermore, the high quality-factor of dark modes can be used to achieve enhanced sensitivity to the dielectric function of the surrounding medium.

Another aspect offered by time-resolved experiments is the possibility to follow energy transfer. In Fig. 5a, the LD resonance in CH$_2$Cl$_2$ broadens in comparison to air. The other SPhPs do not suffer from such a decrease in their lifetime. We apply the same method as previously to extract the lifetime of each mode in CH$_2$Cl$_2$ (Fig. S5). The TQ lifetime changes from 2.2 ± 0.03 ps to 1.3 ± 0.24 ps, which corresponds to a 41 % decrease. For the LD mode, the reduction of the lifetime goes from 4.02 ± 0.13 to 1.17 ± 0.18 ps (~71 % decrease). For each mode, multiple mechanisms explain the decrease, such as additional relaxation channels for phonons in CH$_2$Cl$_2$, chemical interface damping, etc.[47] A possible explanation for the stronger decrease of the monopole is the existence of a vibrational mode of CH$_2$Cl$_2$ with close energy at 8.5 THz. [38] The energetic proximity increases the energy relaxation from the nanowire to the solvent.

Measurements of the lifetime of SPhPs in water also support this conclusion (Fig. S6). In water, each lifetime decreases in a similar fashion, which confirms that the additional decay in $CH_2Cl_2$ is due to the vibrational mode of the solvent. Thanks to the time domain approach we developed, we were able to investigate aspects of vibrational energy transfer. By choosing a solvent with vibrational frequencies overlapping with SPhPs,[19] this method will enable tracking coherence transfer.

We have demonstrated a novel method to study coherent surface phonon polaritons in the time domain. We studied in detail the generation, detection, and nature of these excitations. We then investigated the sensing potential of surface phonon polaritons through FDTD simulations and experiments. We also showed that this technique generates both bright and dark modes. The latter has a long coherence time and will allow more precise sensing. Surface phonon polaritons have quality factors of up to 115, 4 times larger than the highest expected values for plasmonic structures. Finally, access to temporal information allowed us to highlight a pathway for energy transfer from two media through their vibrations. Numerous prospects now become available thanks to the approach we developed. For instance, with the deep confinement of surface phonon polaritons, we unlocked the study of dynamics in the THz range with surface/interface specificity. We also provide an experimental and theoretical toolbox for the quantitative analysis of surface phonon polaritons. Finally, we lay the foundations towards the in-depth time-resolved investigation of the role of coherence in the rich polariton physics.

## Experimental section

**Nanowire growth:**

InP nanowires (NWs) were grown in a low pressure (100 mbar) horizontal metal organic vapor phase epitaxy (MOVPE) system using hydrogen as a carrier gas and a total flow of at 6 slm. 80 nm Au aerosol particles were deposited with an aerial density of 4 m$^{-2}$ onto a (111)B oriented InP:Fe substrate to allow for vapor-liquid-solid NW growth. The InP:Fe substrate was covered by a nominally intrinsic MOVPE-grown InP buffer layer prior to aerosol deposition. Before the actual NW growth, the samples were annealed for 10 minutes at 600 °C in a $PH_3/H_2$ atmosphere to desorb surface oxides. Then the samples were cooled down to the growth temperature of 390 °C, after which growth was initiated by introducing trimethylindium and phosphine at molar fractions of $\chi_{TMIn}= 9.5\times10^{-6}$ and $\chi_{PH3}= 6.2\times10^{-3}$, respectively, for 20 seconds to nucleate the NWs. After the nucleation, the NW growth was continued for 10 min while introducing different molar fractions of $H_2S$ in the range of $H_2S = 0 - 6.3\times10^{-5}$ for the runs. The growth was finished by switching off the TMIn and $H_2S$ supply simultaneously while cooling down to 300 °C in a $PH_3/H_2$ atmosphere.

**Femtosecond pump-probe spectroscopy:**

Experiments were performed using a re-generatively amplified, mode-locked Yb:KGW (Ytterbium-doped potassium gadolinium tungstate) based femtosecond laser system (Pharos, Light conversion) operating at 1030 nm and delivering pulses of 200 fs at 2 kHz repetition rate. This laser is then used to pump two non-collinearly phase-matched optical parametric amplifiers (NOPAs). A first one (Orpheus-N, Light Conversion), was used to generate pump pulses centered at 550 nm with pulse duration of 35 fs. The second NOPA (Orpheus-N, Light Conversion), generated probe pulses at 720 nm with 40 fs pulse duration that were time delayed with respect to the pump. The pump beam was chopped at the frequency of 1 kHz using a mechanical chopper. Both beams were focused on the sample and the modifications of the probe reflectivity induced by the pump were time-resolved. Experiments were performed at normal incidence, and for the polarization dependent study at grazing angle incidence.

**Raman Experiments:**

Raman spectra were acquired under ambient conditions using a high-resolution confocal micro Raman system (Horiba LabRAM HR 800). The laser wavelength used for exciting the samples was 488 nm (2.54 eV).

**FDTD simulations:**

FDTD simulations were performed using the software Lumerical. To model the dielectric properties of the InP nanowires, we used the following dielectric function:

$$\varepsilon = \varepsilon_\infty \left(1 + \frac{\omega_{LO}^2 - \omega_{TO}^2}{\omega_{TO}^2 - \omega^2 + i\omega\gamma}\right)$$

with $\omega_{TO}$ = 8.9 THz, $\omega_{LO}$ = 10.2 THz, the optical phonon frequencies and $1/\gamma$ = 10 ps, the LO phonon lifetime. The damping rate does not influence the frequency of the SPhPs, but plays a role in their lifetime. For both simulations, the sample is composed of an InP nanowire on an InP substrate, and the spectrum of the source spans from 7.5 to 11.5 THz. The diameter and length of nanowires are varied from 90 to 120 nm and 1500 to 2500 nm, respectively, which corresponds to the variations observed in our samples (see Fig. S4). We did not consider the doping of the nanowires as we are interested in the intrinsic modes of the structures. The simulation domain has dimensions of 30x10x10 µm³ with perfectly matching layer boundary conditions. The evolution of the system is monitored for 10 ps. Two types of sources are considered: a plane-wave, which can only couple to bright modes, and a dipole placed in the near-field of the nanowire that couples to both bright and dark modes. In the former case, we calculate the field scattered by the nanowire on substrate using the built-in Total-Field Scattered-Field approach. In the second simulation, we placed a dipole at half the height of the nanowire and 50 nm away from the surface of the wire. We then place a monitor 10 nm above the nanowire and recorded the intensity of the field in this monitor as a function of the frequency, thus revealing the existence of resonant electromagnetic modes of the structure.